\documentclass[11pt]{article}
\usepackage{amssymb}
\usepackage{amsmath}
\usepackage{bbm}
\usepackage{cite}
\usepackage{graphicx}
\usepackage{subfigure}

\newtheorem{theorem}{Theorem}[section]
\newtheorem{prop}[theorem]{Proposition}

\newtheorem{lemma}[theorem]{Lemma}
\newtheorem{define}[theorem]{Definition}

\newcommand\btd{\raise 2pt \hbox{$\hat\bigtriangledown$}\hskip 1.5pt}
\newcommand\bt{\raise 2pt \hbox{$\bigtriangledown$}\hskip 1.5pt}

\def\no{\nonumber}

\begin{document}
\title{An upper order bound of the invariant manifold in Lax pairs of a nonlinear evolution  partial differential equation}
\author{Zhi-Yong Zhang \footnote{E-mail:
zhiyong-2008@163.com} \\
\small~College of Science, Minzu University of China, Beijing 100081, P.R. China}
\date{}
\maketitle
\noindent{\bf Abstract:} In \cite{hab-2016,hab-2017}, Habibullin \emph{et.al} proposed an approach to construct Lax pairs of a nonlinear integrable partial differential equation (PDE), where one is the linearized equation of the studied PDE and the other is the invariant manifold of the linearized equation. In this paper, we show that the invariant manifold is the characteristic of a generalized conditional symmetry of the system composed of the studied PDE and its linearized PDE. Then we give an upper order bound of the invariant manifold which provides a theoretical basis for a complete classification of such type of invariant manifold. As an application, we give a complete classification of the given type invariant manifold for the KdV and mKdV equations and also construct several invariant manifolds and Lax pairs for the Sharma-Tasso-Olver equation.

\section{Introduction}
Lax pair is usually related to the possibility of ``linearizing"
a nonlinear PDE and thus composed of two coupled PDEs linear in an auxiliary dependent variable and compatible on the condition that the considered nonlinear PDE is satisfied \cite{lax}.

Consider a scalar $(1+1)$ dimensional nonlinear evolution PDE in the form
\begin{eqnarray}\label{eqn}
u_t=f(x,t,u,u_1,\dots,u_n), ~~~~n>1
\end{eqnarray}
where $u_0=u$ and $u_j=\partial^j u/\partial x^j, j=1,2,\dots,n$. Introducing the auxiliary variable $\varphi$, Lax pair of Eq.(\ref{eqn}) takes the form
  \begin{equation}\label{lax}
\begin{cases}
\varphi_t=\displaystyle{\sum_{i=0}^{s}\alpha_i(x,t,u,u_1,\dots)}\varphi_i,\\
\varphi_r=\displaystyle{\sum_{i=0}^{r-1}\beta_i(x,t,u,u_1,\dots)}\varphi_i,
\end{cases}
\end{equation}
where $\alpha_i,\beta_i$ are some smooth functions of their arguments, $\varphi_i=\partial^i \varphi/\partial x^i$. In system (\ref{lax}) the first equation is the time-derivative PDE and the second one is the spectral problem, the compatibility condition $\varphi_{tr}=\varphi_{rt}$ holds on the solution manifold of Eq.(\ref{eqn}).

Whether there exist Lax pairs for the considered PDE is a leading tool to discriminate the PDE's integrability, named by Lax integrability. The existence of Lax pair makes us more facilitate to understand the properties of the integrable models. For example, Lax pair is related with infinite conservation laws, and also used to construct B$\ddot{\mbox{a}}$cklund transformation of the PDE. However, Lax pair is not unique and thus many researchers devoted to design effective techniques to find new Lax pairs such as the Zakharov-Shabat dressing \cite{zs-1974} and prolongation structures \cite{we-1975}, etc.

Recently, Habibullin \emph{et.al} suggested a method to construct Lax pairs for the nonlinear integrable equations \cite{hab-2016,hab-2017}. Taking Eq.(\ref{eqn}) as a research object, the first operator in the Lax pair is chosen as the linearized equation of Eq.(\ref{eqn})
\begin{eqnarray}\label{eqn-linear}
v_t=\sum_{i=0}^n \frac{\partial f}{\partial u_i}v_i,~~~v_0=v,
\end{eqnarray}
while the second one is an invariant manifold of Eq.(\ref{eqn-linear}) in the form
\begin{eqnarray}\label{inva-manifold}
H=v_p-\sum_{j=0}^{p-1}\alpha_jv_j=0,
\end{eqnarray}
where $v_j=\partial^j v/\partial x^j$, $\alpha_j=\alpha_j(x,t,u,u_1,\dots,u_s)$ are undetermined smooth functions. Note that the invariant manifold (\ref{inva-manifold}) is not unique and can be taken other expressions such as the quadratic form \cite{hab-2017}. Then Eqs.(\ref{eqn-linear}) and (\ref{inva-manifold}) yield a Lax pair of Eq.(\ref{eqn}) if the following condition holds
\begin{eqnarray}\label{ger-sym-123}
&& D_tH|_{\{(\ref{eqn}),(\ref{eqn-linear}),(\ref{inva-manifold})\}}=0,
\end{eqnarray}
where, hereinafter, $|_{\{\triangle\}}$ means the computations are performed on the solution manifold of $\triangle$, $D_t$ and the subsequent $D_x$ are total differential operators with respect to $t$ and $x$ respectively,
\begin{eqnarray}
&&\no D_t= \frac{\partial }{\partial t}+u_t\frac{\partial }{\partial
u}+u_{1t}\frac{\partial }{\partial u_1}+u_{tt}\frac{\partial
}{\partial u_t}+\cdots,\\
&&\no D_x= \frac{\partial }{\partial x}+u_1 \frac{\partial }{\partial u}+u_{2}\frac{\partial }{\partial u_1}+u_{1t}\frac{\partial }{\partial u_t}
+\cdots.
\end{eqnarray}
Then splitting Eq.(\ref{ger-sym-123}) with respect to $u,v$ and their different order $x$-derivatives gives an over-determined system for $\alpha_j$. Solving the system gives the invariant manifold (\ref{inva-manifold}) and then the required Lax pair is obtained since Eq.(\ref{eqn-linear}) is known.

In this paper, we formulate the method by Habibullin \emph{et.al} into the framework of generalized conditional symmetry and give an upper order bound of $p$ in the invariant manifold (\ref{inva-manifold}), which will promote a complete classification of such type of invariant manifold.  Such results are given in Section 2. In Section 3 we present several examples including the KdV, mKdV and Sharma-Tasso-Olver equations to illustrate the results. The last section concludes the results.

\section{Main results}

\subsection{Related notions}
We first recall some related definitions by considering Eq.(\ref{eqn}).
A generalized symmetry $X=\eta(x,t,u,u_1,\dots)\partial_u$ is admitted by Eq.(\ref{eqn}) if it satisfies
\begin{eqnarray}\label{eqn-sym}
&& X(u_t-f)|_{M}=\bigg(D_t\eta-\sum_{i=0}^n \frac{\partial f}{\partial u_i}D_x^i\eta\bigg)|_{M}=0,
\end{eqnarray}
where, hereinafter, $M$ is the set of all differential consequences of Eq.(\ref{eqn}). Note that on the solution manifold of Eq.(\ref{eqn}) we exclude all derivatives of $u$ with respect to $t$ and thus express $\eta=\eta(x,t,u,u_1,\dots)$. Obviously, Eq.(\ref{eqn-linear}) corresponds to condition (\ref{eqn-sym}) with $\eta=v(x,t)$.

\begin{define}\label{def-1}
(Generalized conditional symmetry \cite{zha-1995,foks-1994}) Eq.(\ref{eqn}) is conditionally invariant under the generalized symmetry $X=\eta(x,t,u,u_1,\dots)\partial_u$ if the following condition
\begin{eqnarray}\label{consymm}
X(u_t-f)|_{M\cap L_x}=0
\end{eqnarray}
holds, where $M$ is defined as in (\ref{eqn-sym}) and $L_x$ denotes the set of all differential consequences of $\eta=0$ with respect to the variable $x$. Then $X$ is called a generalized conditional symmetry of Eq.(\ref{eqn}) and $\eta(x,t,u,u_1,\dots)$ is the characteristic of $X$.
\end{define}

The generalized symmetry is a special case of generalized conditional symmetry. Furthermore, taking $X=\eta(x,t,u,u_1,\dots)\partial_u$  into Eq.(\ref{consymm}) yields
\begin{eqnarray}
\no X(u_t-f)|_{M\cap L_x}=\bigg(D_t\eta-\sum_{i=0}^n \frac{\partial f}{\partial u_i}D_x^i\eta\bigg)|_{M\cap L_x}=D_t\eta|_{M\cap L_x}=0,
\end{eqnarray}
which is a simple condition to determine generalized conditional symmetry of Eq.(\ref{eqn}). Let $L$ denotes the set of all differential results of $\eta=0$, then the solution manifold $M\cap L$ is contained in $M\cap L_x$, thus
\begin{eqnarray}\label{con-man}
D_t\eta|_{M\cap L}=0.
\end{eqnarray}

\begin{define}\label{def-2}(Invariant manifold \cite{kap-1992})
An ordinary differential equation
\begin{eqnarray}\label{inv-set}
I(x,t,u,u_1,\dots,u_m)=0
\end{eqnarray}
defines an invariant manifold of Eq.(\ref{eqn}) if it satisfies
\begin{eqnarray}\label{inv-set-con}
D_tI(x,t,u,u_1,\dots,u_m)|_{\{(\ref{eqn}),(\ref{inv-set})\}}=0.
\end{eqnarray}
\end{define}

By Definition \ref{def-1} and equality (\ref{con-man}), condition (\ref{inv-set-con}) means that $X=I(x,t,u,u_1,\dots,u_m)\partial_u$ with $I$ given in (\ref{inv-set}) is a generalized conditional symmetry of Eq.(\ref{eqn}). Thus the system
 \begin{equation}\no
\begin{cases}
u_t=f(x,t,u,u_1,\dots,u_n),\\
I(x,t,u,u_1,\dots,u_m)=0,
\end{cases}
\end{equation}
is compatible \cite{zha-1995}.

\subsection{Construction of Lax pair}
Consider a general form invariant manifold of Eq.(\ref{eqn-linear})
\begin{eqnarray}\label{inv-set-special}
&& H(x,t,u,u_1,\dots,u_m;v,v_1,\dots,v_p)=0,
\end{eqnarray}
by the method in \cite{hab-2016}, it means
\begin{eqnarray}\label{ger-sym-1}
&& D_tH(x,t,u,u_1,\dots,u_m;v,v_1,\dots,v_p)|_{\{(\ref{eqn}),(\ref{eqn-linear}),(\ref{inv-set-special})\}}=0.
\end{eqnarray}

\begin{lemma}\label{th-3}
Following the above notations and conditions, Eqs.(\ref{eqn-linear}) and (\ref{inv-set-special}) are consistent on the solution space of Eq.(\ref{eqn}).
\end{lemma}

\emph{Proof:} Since Eq.(\ref{inv-set-special}) is an invariant manifold of Eq.(\ref{eqn-linear}), then by Definition \ref{def-1} and condition (\ref{ger-sym-1}), we get that $X=H\partial_v$ with $H$ defined by (\ref{inv-set-special}) is a generalized conditional symmetry of Eqs.(\ref{eqn}) and (\ref{eqn-linear}),
thus Eqs.(\ref{eqn-linear}) and (\ref{inv-set-special}) are consistent on the solution space of Eq.(\ref{eqn}) \cite{zha-1995}. The proof ends. $\hfill{} \Box$

In particular, if the invariant manifold takes the special form (\ref{inva-manifold}), then by Lemma \ref{th-3}, Eqs.(\ref{eqn-linear}) and (\ref{inva-manifold}) are consistent and their compatibility generates Eq.(\ref{eqn}) and thus can be regarded as a Lax pair of  Eq.(\ref{eqn}).

\begin{lemma}\label{th-1}
Let $X=H\partial_v$ with $H$ given in (\ref{inv-set-special}) is a generalized conditional symmetry of Eqs.(\ref{eqn}) and (\ref{eqn-linear}), then $H=0$ defines an invariant manifold of  Eq.(\ref{eqn-linear}).
\end{lemma}

\emph{Proof:} Since $X=H\partial_v$ with $H$ given in (\ref{inv-set-special}) is a generalized conditional symmetry of Eqs.(\ref{eqn}) and (\ref{eqn-linear}), then
\begin{eqnarray}\label{ger-sym}
&&\no X(u_t-f)|_{\{(\ref{eqn}),(\ref{eqn-linear}),(\ref{inv-set-special})\}}=0,\\
&&  X\bigg(v_t-\sum_{i=0}^n \frac{\partial f}{\partial u_i}v_i\bigg)|_{\{(\ref{eqn}),(\ref{eqn-linear}),(\ref{inv-set-special})\}}=0
\end{eqnarray}

Obviously, the first equation in  system (\ref{ger-sym}) holds identically since it is independent of $v$ and its derivatives, while inserting $X=H\partial_v$ into the second equation yields
\begin{eqnarray}
&&\no X\bigg(v_t-\sum_{i=0}^n \frac{\partial f}{\partial u_i}v_i\bigg)|_{\{(\ref{eqn}),(\ref{eqn-linear}),(\ref{inv-set-special})\}}=\bigg(D_tH-\sum_{i=0}^n \frac{\partial f}{\partial u_i}D_x^iH\bigg)|_{\{(\ref{eqn}),(\ref{eqn-linear}),(\ref{inv-set-special})\}}\\\no
&&\hspace{5.3cm}=D_tH|_{\{(\ref{eqn}),(\ref{eqn-linear}),(\ref{inv-set-special})\}}\\\no
&&\hspace{5.3cm}=0,
\end{eqnarray}
which is the same to condition (\ref{ger-sym-1}). It completes the proof.$\hfill{} \Box$

Note that alternatively one can choose $X=H\partial_u$ as a generalized conditional symmetry of Eqs.(\ref{eqn}) and (\ref{eqn-linear}), then with similar proof we find that $H=0$ also defines an invariant manifold of  Eq.(\ref{eqn-linear}). Moreover, one can also call $X$ as a generalized conditional symmetry of Eq.(\ref{eqn-linear}) with the nonlinear differential constraint Eq.(\ref{eqn}), but here we adopt the former statement.

Therefore, by Lemma \ref{th-1} a simple and practical technique to find invariant manifold of Eq.(\ref{eqn-linear}) is to search for a generalized conditional symmetry $X=H\partial_v$ of the system consisting of Eq.(\ref{eqn}) and Eq.(\ref{eqn-linear}). Furthermore, the existence of the invariant manifold in the form (\ref{inva-manifold}) is guaranteed by the following result.
\begin{lemma}
Eqs.(\ref{eqn}) and (\ref{eqn-linear}) always have the invariant manifold of the form (\ref{inva-manifold}).
\end{lemma}

\emph{Proof. } Observe that Eqs.(\ref{eqn}) and (\ref{eqn-linear}) are always admitted by the symmetry $X_v=v\partial_v$ since on the solution space of Eqs.(\ref{eqn}) and (\ref{eqn-linear}),
 \begin{equation}\label{sys-1}
\begin{cases}
 X_v(u_t-f)\equiv0,\\
\displaystyle{\mbox{Pr}^{(n)}X_v\bigg(v_t-\sum_{i=0}^n \frac{\partial f}{\partial u_i}v_i\bigg)|_{\{(\ref{eqn}),(\ref{eqn-linear})\}}=\bigg(v_t-\sum_{i=0}^n \frac{\partial f}{\partial u_i}v_i\bigg)}|_{\{(\ref{eqn}),(\ref{eqn-linear})\}}=0,
\end{cases}
\end{equation}
where $\mbox{Pr}^{(n)}X_v$ denotes the $n$th-order prolongation of $X_v$ and is given by $\mbox{Pr}^{(n)}X_v=v_t\partial_{v_t}+\sum_{i=0}^n v_i\partial_{v_i}$ with $v_0=v$ \cite{olv}. It has $n$ differential variants $v_i/v\,(i=1,\dots,n)$. Thus the invariant manifold $H=0$ spanned by the linear combinations of $v_i/v\,(i=1,\dots,p)$ takes the form (\ref{inva-manifold}). The proof ends. $\hfill{} \Box$

Note that if $p<n$ then the invariant manifold is spanned by $p$ differential invariants $v_i/v\,(i=1,\dots,p)$, otherwise one can prolong $X$ to $p$ th-order, $\mbox{Pr}^{(p)}X_v=v_t\partial_{v_t}+\sum_{i=0}^p v_i\partial_{v_i}$, which also makes system (\ref{sys-1}) hold, then the differential variants are $v_i/v\,(i=1,\dots,p)$.

Summarizing the above Lemmas, we give the following theorem to determine the invariant manifold of Eq.(\ref{eqn-linear}).
\begin{theorem}\label{th-2}
The invariant manifold of Eq.(\ref{eqn-linear}) with the form (\ref{inva-manifold}) is the characteristic of a generalized conditional symmetry of Eq.(\ref{eqn}) and Eq.(\ref{eqn-linear}),  determined by $D_tH|_{\{(\ref{eqn}),(\ref{eqn-linear}),(\ref{inva-manifold})\}}=0$.
\end{theorem}

For example, consider the well-known KdV equation
\begin{eqnarray}\label{kdv}
&& u_t=u_{xxx}+uu_x.
\end{eqnarray}
Its linearized equation is
\begin{eqnarray}\label{kdv-linearl}
&& v_t=v_{xxx}+uv_x+u_xv,
\end{eqnarray}
which admits a linear invariant manifold with $p=3$ \cite{hab-2016}
\begin{eqnarray}\label{kdv-inv}
&&\no 0=H=v_3-\frac{u_2}{u_1}v_2+\bigg(\frac{2}{3}u+\lambda\bigg)v_1-\bigg(\frac{2 uu_2}{3u_1}+\lambda\frac{u_2}{u_1}- u_1\bigg)v\\
&& \hspace{1.16cm}=v\left[\frac{v_3}{v}-\frac{u_2}{u_1}\frac{v_2}{v}+\bigg(\frac{2}{3}u+\lambda\bigg)\frac{v_1}{v}-\bigg(\frac{2 uu_2}{3u_1}+\lambda\frac{u_2}{u_1}- u_1\bigg)\right].
\end{eqnarray}

Actually, let $X=H\partial_v$ with $H$ given in (\ref{kdv-inv}), then on the solution manifold of Eqs.(\ref{kdv}) and (\ref{kdv-linearl}), we find
\begin{eqnarray}
&&\no \mbox{Pr}^{(3)}X(u_t-uu_x-u_{xxx})\equiv0,
\end{eqnarray}
while direct computations yield
\begin{eqnarray}
&&\no \mbox{Pr}^{(3)}X\left(v_t-uv_x-u_xv-v_{xxx}\right)|_{\{(\ref{kdv}),(\ref{kdv-linearl}),(\ref{kdv-inv})\}}=\left(D_tH-u D_xH-u_x H-D_x^3H\right)|_{\{(\ref{kdv}),(\ref{kdv-linearl}),(\ref{kdv-inv})\}}\\
&&\no \hspace{7.5cm}=D_tH|_{\{(\ref{kdv}),(\ref{kdv-linearl}),(\ref{kdv-inv})\}}\\
&&\no \hspace{7.5cm}=0.
\end{eqnarray}

Therefore, $X=H\partial_v$ with $H$ given in (\ref{kdv-inv}) is a generalized conditional symmetry of the system consisting of Eqs.(\ref{kdv}) and (\ref{kdv-linearl}).

\subsection{An upper order bound of the invariant manifold}
By extending the idea of maximal dimension of invariant subspace of nonlinear PDE \cite{vas-2007,qu-2014}, we give an upper order bound of the invariant manifold (\ref{inva-manifold}) of Eq.(\ref{eqn-linear}).
\begin{theorem}\label{the-2}
Let the derivatives of highest-order $s$ in $\alpha_j$ satisfying $s\leq n$. Then the order $p$ in Eq.(\ref{inva-manifold}) is bounded by $p\leq 2n+1$, where $n$ is the order of nonlinear Eq.(\ref{eqn}).
\end{theorem}

\emph{Proof:} We prove it by contradiction and thus assume $p> 2n+1$ to show Eq.(\ref{eqn}) is linear.
By the definition of invariant manifold, on the solution manifold of the system composed of equations (\ref{eqn}), (\ref{eqn-linear}) and (\ref{inva-manifold}), one has
\begin{eqnarray}\label{th-part}
&& D_tH=D_t\bigg(v_p-\sum_{j=0}^{p-1}\alpha_jv_j\bigg)=D_t(v_p)-D_t\sum_{j=0}^{p-1}\big(\alpha_jv_j\big)=0.
\end{eqnarray}

Observe that keeping the highest order derivative in $u$ yields
\begin{eqnarray}
&&\no D_x\frac{\partial f}{\partial u_n}=\frac{\partial^2 f}{\partial u_n^2}u_{n+1}+\mbox{the terms with order in $u$ less than}\, (n+1),\\
&&\no D_x^2\frac{\partial f}{\partial u_n}=\frac{\partial^2 f}{\partial u_n^2}u_{n+2}+\mbox{the terms with order in $u$ less than}\, (n+2),\\
&&\no\hspace{5cm} \dots,\\
&&\no D_x^l\frac{\partial f}{\partial u_n}=\frac{\partial^2 f}{\partial u_n^2}u_{n+l}+\mbox{the terms with order in $u$ less than}\, (n+l).
\end{eqnarray}

For the term $D_t(v_p)$ in Eq.(\ref{th-part}), we express all the derivatives of $u_t$ by Eq.(\ref{eqn}), $v_t$ by Eq.(\ref{eqn-linear}) and $v_p$ by Eq.(\ref{inva-manifold}) respectively, then we get
\begin{eqnarray}\label{th-part-1}
&&\no D_t(v_p)=\sum_{i=0}^n D_x^p\bigg(\frac{\partial f}{\partial u_i}v_i\bigg)\\\no
%&&\hspace{0.9cm}=v_p \bigg(\frac{\partial \alpha}{\partial t}+\sum_{k=0}^{s}\frac{\partial\alpha_p}{\partial u_k}D_x^k f\bigg)+\alpha_p\sum_{i=0}^n \sum_{l=0}^p C_p^lv_{i+p-l} D_x^l\bigg(\frac{\partial f}{\partial u_i}\bigg)\\\no
&&\hspace{1.2cm}=\sum_{l=0}^p C_p^lv_{n+p-l} D_x^l\bigg(\frac{\partial f}{\partial u_n}\bigg)+\sum_{i=0}^{n-1} \sum_{l=0}^p C_p^lv_{i+p-l} D_x^l\bigg(\frac{\partial f}{\partial u_i}\bigg)\\
&&\hspace{1.2cm}=v_{n+p}\frac{\partial f}{\partial u_n}+\sum_{l=1}^p C_p^lv_{n+p-l}  u_{l+n}\bigg(\frac{\partial^2 f}{\partial u_n^2}\bigg)+\dots,
%&&\hspace{1.9cm}+\alpha_p\sum_{l=1}^p \sum_{s=0}^{n-1} C_p^lv_{n+p-l} u_{l+s}\Big(\frac{\partial^2 f}{\partial u_s\partial u_n}\Big)+\alpha_p\sum_{s=0}^{n-1} \sum_{l=1}^p C_p^lv_{s+p-l} u_{n+l}\Big(\frac{\partial^2 f}{\partial u_s\partial u_n}\Big)+\dots\\\no
%&&\hspace{1.9cm}+\sum_{l=1}^p \sum_{s=0}^{n-1} C_p^l\left(v_{n+p-l} u_{l+s}+v_{s+p-l} u_{n+l}\right)\Big(\frac{\partial^2 f}{\partial u_s\partial u_n}\Big)+\dots\\\no
%&&\hspace{0.9cm}=v_{n+p}\frac{\partial f}{\partial u_n}+u_{n+p}\sum_{i=0}^n \Big(\frac{\partial^2 f}{\partial u_i\partial u_n}v_i\Big)+\sum_{i=1}^{p-1} \big(C_p^i\,v_{n+p-i}u_{n+i}\big)\,\frac{\partial^2 f}{\partial u_n^2}\\
%&&\hspace{1.5cm}+\bigg[\sum_{i=1}^{[\frac{p}{2}]-1} C_p^i\,u_{n+p-i}u_{n+i}+\gamma C_p^{[\frac{p}{2}]}u_{n+p-[\frac{p}{2}]}u_{n+[\frac{p}{2}]}\bigg]\sum_{i=0}^n \Big(\frac{\partial^3 f}{\partial u_i\partial u_n^2}v_i\Big)+\dots,
\end{eqnarray}
where we display the highest order derivative term $v_{n+p}$ and the quadratic term with maximal total order in derivatives of $u$ and $v$.
Notice that the maximal total order in the quadratic terms is $(n+p-l)+(n+l)=2n+p$. %In particular, consider the quadratic term in Eq.(\ref{th-part-1})

Consider the quadratic term $\sum_{l=1}^{p}\big( C_p^l\,v_{n+p-l}u_{n+l}\big)\partial^2 f/\partial u_n^2$ in (\ref{th-part-1}). Since $p\geq2n+2$, we select the derivatives in $v$ of the order not less than $p-1$,
\begin{eqnarray}\label{qua-1}
&& \sum_{l=1}^{n+1}\big( C_p^l\,v_{n+p-l}u_{n+l}\big)\,\frac{\partial^2 f}{\partial u_n^2}.
\end{eqnarray}

Then by Eq.(\ref{inva-manifold}), we express all the derivatives $v_{n+p-l}$ for $l=1,2,\dots,n$ in terms of $v_{p-1},\dots,v$ and isolate the terms involving $v_{p-1}$ in (\ref{qua-1}),
\begin{eqnarray}\label{qua-2}
&& \no\left[C_p^{n+1}u_{2n+1}+\sum_{l=1}^{n}\big(\widetilde{ \alpha}_l u_{n+l}\big)\right] v_{p-1}\,\frac{\partial^2 f}{\partial u_n^2}+\dots,
\end{eqnarray}
where $\widetilde{\alpha}_l$ are expressed via $\alpha_i$ in (\ref{inva-manifold}) and their derivatives.
Then (\ref{qua-1}) will contain a single quadratic term $C_p^{n+1} v_{p-1}u_{2n+1}\,\partial^2 f/\partial u_n^2$ which has the maximal total order $2n+p$.

Similarly, the second part of the last equality in (\ref{th-part}) can be expressed as
\begin{eqnarray}\label{th-part-2}
&&\no D_t\Bigg(\sum_{j=0}^{p-1}\alpha_jv_j\Bigg)=\sum_{j=0}^{p-1}\big(v_j\, D_t\alpha_j+\alpha_j\,v_{jt}\big)\\\no
%&&\hspace{2.6cm}=\sum_{j=0}^{p-1}\bigg[v_j\, D_t\alpha_j+\alpha_j\,D_x^j\Big(\sum_{i=0}^n \frac{\partial^i f}{\partial u_i}v_i\Big)\bigg]\\
%&&\hspace{2.6cm}=\sum_{j=0}^{p-1}\bigg[v_j\, D_t\alpha_j+\alpha_j\sum_{i=0}^n D_x^j\Big(\frac{\partial^i f}{\partial u_i}v_i\Big)\bigg]\\
&&\hspace{2.6cm}=\sum_{j=0}^{p-1}\left[v_j \bigg(\frac{\partial \alpha_j}{\partial t}+\sum_{k=0}^{s}\frac{\partial\alpha_j}{\partial u_k}u_{kt}\bigg)+\alpha_j\sum_{i=0}^n D_x^j\bigg(\frac{\partial f}{\partial u_i}v_i\bigg)\right]\\
&&\hspace{2.6cm}=\sum_{j=0}^{p-1}\left[v_j \bigg(\frac{\partial \alpha_j}{\partial t}+\sum_{k=0}^{s}\frac{\partial\alpha_j}{\partial u_k}D_x^k f\bigg)\right]+\sum_{j=0}^{p-1}\left[\alpha_j\sum_{i=0}^n D_x^j\bigg(\frac{\partial f}{\partial u_i}v_i\bigg)\right],
\end{eqnarray}
where the maximal total order of quadratic terms in derivatives of $u$ and $v$ is $n+s+p-1 \leq 2n+p-1 $ since $s\leq n$. Thus by considering Eq.(\ref{th-part}), the terms $C_p^{n+1} v_{p-1}u_{2n+1}\,\partial^2 f/\partial u_n^2$ with the total order $2n+p$ in $D_t(v_p)$ is unique and must be vanished, thus
\begin{eqnarray}
&&\no \frac{\partial^2 f}{\partial u_n^2}=0,
\end{eqnarray}
which means $f=\lambda_n(x,t,u,\dots,u_{n-1})u_n+\widetilde{f}(x,t,u,u,u_1,\dots,u_{n-1})$.

Taking the expression of $f$ into account, similar as (\ref{qua-1}), we find
\begin{eqnarray}\label{th-part-11}
&&\no D_t(v_p)=D_x^p(v_t)
=\sum_{i=1}^{n+1} \big(\beta_i v_{n+p-i}u_{n-1+i}\big)\,\frac{\partial^2 f}{\partial u_n\partial u_{n-1}}+\dots,
\end{eqnarray}
where $\beta_i>0$.  All the summands in quadratic terms have the total order of the derivatives $2n+p-1$. Excluding the derivatives $v_{n+p-i}$ for $i=1,2,\dots,n$ by Eq.(\ref{inva-manifold}) gives the unique quadratic term $$\beta_{n+1}v_{p-1}u_{2n}\frac{\partial^2 f}{\partial u_n\partial u_{n-1}}, $$ which cannot occur in Eq.(\ref{th-part-2}), thus $\partial^2 f/\partial u_n\partial u_{n-1}=0$. Similarly, by induction, we obtain
\begin{eqnarray}
&&\no \frac{\partial^2 f}{\partial u_n\partial u_i}=0,~~~i=0,1,\dots, n-2,
\end{eqnarray}
which implies that  $f=\beta_n(x,t)u_n+\widetilde{f}(x,t,u,u,u_1,\dots,u_{n-1})$, i.e., $f$ depends on $u_n$ linearly.

Repeating the same procedures for the function $\widetilde{f}$ yields $\widetilde{f}=\beta_{n-1}(x,t)u_{n-1}+\widetilde{\widetilde{f}}(x,t,u,u,u_1,$ $u_2,\dots,u_{n-2})$, and finally, we find
$$f=\sum_{i=0}^{n}\beta_i(x,t)u_i+\beta_{n+1}(x,t),$$
which means Eq.(\ref{eqn}) is linear, it contradicts. The proof ends. $\hfill{} \Box$

It should be noted that in \cite{hab-2018a,hab-2018b} the authors stated that any set of symmetries of Eq.(\ref{eqn}) defines an invariant manifold (\ref{inv-set-special}) of Eq.(\ref{eqn-linear}) and built a connection with the recursion operator, thus it seems that the order of the invariant manifold (\ref{inv-set-special}) is infinite. However, the invariant manifold (\ref{inva-manifold}) in the Lax pair of Eq.(\ref{eqn}) is linear in $v_i$ and with the assumption that the order in $\alpha_i$ is bounded by $s\leq n$, then the order bound in Theorem \ref{the-2} is suitable for the invariant manifold with the form (\ref{inva-manifold}).

Once the maximal dimension of the invariant manifold(\ref{inva-manifold}) is estimated, then we can use the invariant manifold linear in $v_i$ up to the maximal order to construct Lax pairs of Eq.(\ref{eqn}). In particular, if the invariant manifold $H=0$ takes the form (\ref{inva-manifold}), then the condition  $D_tH|_{\{(\ref{eqn}),(\ref{eqn-linear}),(\ref{inva-manifold})\}}=0$ is equivalent to
\begin{eqnarray}
\sum_{i=0}^n D_x^p\left(\frac{\partial f}{\partial u_i}v_i\right)_{|_{\{(\ref{inva-manifold})\}}}-\sum_{j=0}^{p-1}\left[v_j \sum_{k=0}^{s}\Big(\frac{\partial\alpha_j}{\partial u_k}D_x^k f\Big)+\alpha_j\sum_{i=0}^n D_x^j\Big(\frac{\partial f}{\partial u_i}v_i\Big)\right]_{|_{\{(\ref{inva-manifold})\}}}=0.
\end{eqnarray}

\section{An algorithm and applications}\label{section}
Following the above analysis, we first state an algorithm to classify the given type of Lax pair of Eq.(\ref{eqn}) and then apply the algorithm to the KdV, mKdV and Sharma-Tasso-Olver equations. %consider several examples by means of the algorithm.

The Lax pair composed of Eq.(\ref{eqn-linear}) and (\ref{inva-manifold}) for Eq.(\ref{eqn}) is constructed by the following steps:

1). Write down the linearized equation of Eq.(\ref{eqn}) and assume the time-derivative PDE in the Lax pair takes the form
\begin{eqnarray}\label{eqn-linear-new}
&& v_t=\sum_{i=0}^n \frac{\partial f}{\partial u_i}v_i,~~~v_0=v.
\end{eqnarray}

2). Suppose the spectral problem in the Lax pair to be the form (\ref{inva-manifold}) and then determine the coefficients $\alpha_j$ via the condition  $D_tH|_{\{(\ref{eqn}),(\ref{eqn-linear}),(\ref{inva-manifold})\}}=0$, where the differential order bound of $v$ is $p\leq 2n+1$ by Theorem \ref{the-2}.

This step is equivalent to search for an invariant manifold of Eq.(\ref{eqn-linear-new}) which takes the form (\ref{inva-manifold}) and contains a constant parameter.

3). Collecting Eqs.(\ref{eqn-linear-new}) and (\ref{inva-manifold}) determined in Step 2) yields a Lax pair of Eq.(\ref{eqn}).

\subsection{The KdV equation}

We consider the well-known KdV equation written in the form
 \begin{equation}\label{kdv-new}
G=u_t-uu_x-u_{xxx}=0,
\end{equation}
whose linearized equation is
 \begin{equation}\label{kdv-linear}
G^L=v_t-u_xv-uv_x-v_{xxx}=0.
\end{equation}

We seek the linear invariant manifold with the form
 \begin{equation}\label{kdv-inva}
H^p=v_p-\sum_{j=0}^{p-1}\alpha_jv_j=0,
\end{equation}
where, hereinafter, $\alpha_j=\alpha_j(u,u_x,u_{xx},u_{xxx})$ and $p\leq 7$ by Theorem \ref{the-2}. In particular, the case for $p=2$ has been considered in \cite{hab-2016}, which demonstrates that for $p=2$ there does not exist the invariant manifold $v_2=\alpha_1(u,u_x,u_{xx})v_1+\alpha_0(u,u_x,u_{xx})v$. The case for $p=3$ has been considered in  \cite{hab-2016} and yields (\ref{kdv-inv}), thus we proceed from $p=4$.
\begin{prop}\label{prop-1}
For $p=4$, the invariant manifold is
 \begin{equation}\label{in-p=4}
H^4=v_4-\frac{u_3}{u_2}v_3+\left(\frac{2}{3}u-\lambda\right)v_2+\left[\frac{5}{3}u_1+\frac{u_3}{u_2}\left(\lambda-\frac{2}{3}u\right)\right]v_1
-\left(\frac{u_1u_3}{u_2}-\frac{4}{3}u_2\right)v=0.
\end{equation}
For $5\leq p\leq7$, there are no invariant manifolds with the form (\ref{kdv-inva}) for Eq.(\ref{kdv-new}).
\end{prop}

\emph{Proof.}
Substituting $H^p$ given in (\ref{kdv-inva}) and $f=uu_1+u_{3}$ into $D_t(H^p)|_{\{(\ref{kdv-new}),(\ref{kdv-linear}), (\ref{kdv-inva})\}}=0$ gives
 \begin{eqnarray}\label{kdv-inva-kp}
&&\no D_t(H^p)=D_t(v_p-\sum_{j=0}^{p-1}\alpha_jv_j)\\
&&\no \hspace{1.4cm}=D_x^p\left(u_1v+uv_1+v_{3}\right)-\sum_{j=0}^{p-1}\left[v_j(D_t\alpha_j)+\alpha_j(D_t v_j)\right]\\
&& \hspace{1.4cm}=\Bigg(D_x^p-\sum_{j=0}^{p-1}\alpha_jD_x^j \Bigg)\left(u_1v+uv_1+v_{3}\right)-\sum_{j=0}^{p-1}\sum_{k=0}^{3}v_j\frac{\partial\alpha_j}{\partial u_k}D_x^k (uu_1+u_{3}),
\end{eqnarray}
which holds on the solution manifold of Eq.(\ref{kdv-inva}).
By separating Eq.(\ref{kdv-inva-kp}) with respect to $u_j\,(4\leq j\leq p+1)$, $v$ and their $x$-derivatives, we get an over-determined system for $\alpha_j$ and then solve it to find  $\alpha_j$.

We consider the highest order derivatives of $u$ in Eq.(\ref{kdv-inva-kp}) with $4\leq p\leq7$ and divide the value of $p$ into two cases, $p=4$ and $5\leq p\leq7$. For the latter case, the highest order derivative term in Eq.(\ref{kdv-inva-kp}) appears in $D_x^p(u_1v)$ and is $u_{p+1}v$ whose coefficient is $1$, thus the corresponding determining system has no solutions.

For $p=4$, on the solution manifold of $H^4=0$, Eq.(\ref{kdv-inva-kp}) becomes
 \begin{eqnarray}\label{kdv-inva-kp1}
&& D_t(H^4)=\left(D_x^4-\sum_{j=0}^3\alpha_jD_x^j \right)\left(u_1v+uv_1+v_{3}\right)-\sum_{j=0}^3\sum_{k=0}^{3}v_j\frac{\partial\alpha_j}{\partial u_k}D_x^k (uu_1+u_{3}).
\end{eqnarray}

Observe that the terms involving $u_6$ uniquely appear in $D_x^4(v_3)$ and $\sum_{j=0}^3 \partial\alpha_j/\partial u_3\,u_{6}v_j$ and cancel out since \begin{eqnarray}
&&\no D_x^4(v_{3})=D_x^3(\sum_{j=0}^{3}\alpha_jv_j)=\sum_{j=0}^3 \frac{\partial\alpha_j}{\partial u_3}\,u_{6}v_j+\mbox{the terms with order in $u$ less than}\, (n+1),
\end{eqnarray}
 thus we consider the terms containing $u_5$. In particular, the coefficients of $u_5u_4 v_i$ with $i=0,1,2,3$ gives $\partial^2\alpha_i/\partial u_i^2=0$ which means $\alpha_i=\alpha_{i1}(u,u_1,u_2) u_3+\alpha_{i2}(u,u_1,u_2)$ respectively. Then substituting $\alpha_i$ into Eq.(\ref{kdv-inva-kp1}) and reconsidering the coefficient of $u_5v_3$ yields
 \begin{eqnarray}\label{kdv-inva-kp2}
&&\alpha_{21}+\alpha_{32}\alpha_{31}+u_3\left(\alpha_{31}^2+\frac{\partial \alpha_{31}}{\partial u_2}\right)+u_2\frac{\partial \alpha_{31}}{\partial u_1}+u_1\frac{\partial \alpha_{31}}{\partial u}+\frac{\partial \alpha_{31}}{\partial x}=0.
\end{eqnarray}
Since $\alpha_{31}$ is independent of $u_3$, further separation of Eq.(\ref{kdv-inva-kp2}) gives
 \begin{eqnarray}\label{kdv-inva-kp2-1}
&&\no \alpha_{31}^2+\frac{\partial \alpha_{31}}{\partial u_2}=0\\
&&\alpha_{21}+\alpha_{32}\alpha_{31}+u_2\frac{\partial \alpha_{31}}{\partial u_1}+u_1\frac{\partial \alpha_{31}}{\partial u}+\frac{\partial \alpha_{31}}{\partial x}=0.
\end{eqnarray}
Similarly, the coefficients of $u_5v_i$ with $i=0,1,2$ are
\begin{eqnarray}\label{kdv-inva-kp3}
&&\no \alpha_{i1}\alpha_{31}+\frac{\partial \alpha_{i1}}{\partial u_2}=0,~~~i=0,1,2,\\
&&\no \frac{1}{3}+\alpha_{02}\alpha_{31}+u_2\frac{\partial \alpha_{01}}{\partial u_1}+u_1\frac{\partial \alpha_{01}}{\partial u}+\frac{\partial \alpha_{01}}{\partial x}=0,\\
&&\no \alpha_{01}+\alpha_{12}\alpha_{31}+u_2\frac{\partial \alpha_{11}}{\partial u_1}+u_1\frac{\partial \alpha_{11}}{\partial u}+\frac{\partial \alpha_{11}}{\partial x}=0,\\
&& \alpha_{11}+\alpha_{22}\alpha_{31}+u_2\frac{\partial \alpha_{21}}{\partial u_1}+u_1\frac{\partial \alpha_{21}}{\partial u}+\frac{\partial \alpha_{21}}{\partial x}=0.
\end{eqnarray}
Solving systems (\ref{kdv-inva-kp2-1}) and (\ref{kdv-inva-kp3}) gives
\begin{eqnarray}\label{equa-3}
&&\no \alpha_{31}=\frac{1}{u_2-\beta_3},~~\alpha_{i1}=\frac{\beta_i}{\beta_3-u_2},~~i=0,1,2,\\
&&\no \alpha_{32}=\beta_2+\frac{1}{u_2-\beta_3}\left(\frac{\partial \beta_3}{\partial u}u_1+\frac{\partial \beta_3}{\partial u_1}u_2\right),\\
&&\no \alpha_{22}=\beta_1+\frac{\partial \beta_2}{\partial u}u_1+\frac{\partial \beta_2}{\partial u_1}u_2-\frac{\beta_2}{u_2-\beta_3}\left(\frac{\partial \beta_3}{\partial u}u_1+\frac{\partial \beta_3}{\partial u_1}u_2\right),\\
&&\no \alpha_{12}=\beta_0+\frac{\partial \beta_1}{\partial u}u_1+\frac{\partial \beta_1}{\partial u_1}u_2-\frac{\beta_1}{u_2-\beta_3}\left(\frac{\partial \beta_3}{\partial u}u_1+\frac{\partial \beta_3}{\partial u_1}u_2\right),\\
&& \alpha_{02}=\frac{1}{3}\beta_3+\beta_0\frac{\partial \beta_3}{\partial u_1}+\frac{\partial \beta_0}{\partial u}u_1+\left(\frac{\partial \beta_0}{\partial u_1}-\frac{1}{3}\right)u_2+\frac{\beta_0}{u_2-\beta_3}\left(\frac{\partial \beta_3}{\partial u}u_1+\beta_3\frac{\partial \beta_3}{\partial u_1}\right),
\end{eqnarray}
where $\beta_j=\beta_j(u,u_1)$ with $j=0,1,2,3$.

Inserting such results into Eq.(\ref{kdv-inva-kp1}) and extracting the coefficient of $u_4v_3u_2^2$ gives
$2\partial\beta_2/\partial u_1-\partial^2\beta_3/\partial u_1^2=0$
while the coefficients of $u_4v_iu_2^3$ gives $\partial^2\beta_i/\partial u_1^2=0$ with $i=0,1,2$. Then by solving them we obtain
\begin{eqnarray}
&&\no \beta_i=\beta_{i1}(u)u_1+\beta_{i2}(u),~~~~~~i=0,1,2,\\
&& \beta_3=\beta_{21}(u)u_1^2+\beta_{32}(u)u_1+\beta_{33}(u).
\end{eqnarray}

Again substituting them into Eq.(\ref{kdv-inva-kp1}) and extracting the coefficient of $u_4v_3u_2$ yields
\begin{eqnarray}
\left(\frac{\partial\beta_{21}}{\partial u}+
   \beta_{21}^2\right) u_1^2+\left( \beta_{21}\beta_{32}-\frac{\partial\beta_{22}}{\partial u}+ \frac{\partial\beta_{32}}{\partial u}\right)u_1+\beta_{21}\beta_{33}=0,
\end{eqnarray}
which gives by further separation with respect to $u_1$
\begin{eqnarray}\label{equa-1}
&& \beta_{21}\beta_{33}=0, ~~~\frac{\partial\beta_{21}}{\partial u}+ \beta_{21}^2=0,~~~\beta_{21}\beta_{32}-\frac{\partial\beta_{22}}{\partial u}+ \frac{\partial\beta_{32}}{\partial u}=0.
\end{eqnarray}
Following similar procedure for the coefficients of $u_4v_iu_2^2$ with $i=0,1,2$, we obtain
\begin{eqnarray}\label{equa-2}
&&\no \beta_{11}+\beta_{21}\beta_{22}=0,\\
&&\no 1+\beta_{01}+\beta_{12}\beta_{21}=\beta_{11}\beta_{21}+\frac{\partial\beta_{11}}{\partial u}=0,\\
&&3\beta_{02}\beta_{21}-\beta_{22}+\beta_{32}=\beta_{21}+3\beta_{01}\beta_{21}+3\frac{\partial\beta_{01}}{\partial u}=0.
\end{eqnarray}

We claim that $\beta_{33}=0$. If $\beta_{33}\neq0$, then $\beta_{21}=0$ and from system (\ref{equa-2}), we find $\beta_{11}=0,\beta_{01}=-1,\beta_{32}=\beta_{22}$. Then the determining equations involve two equations $$1+\frac{\partial\beta_{12}}{\partial u}+\beta_{22}\frac{\partial\beta_{22}}{\partial u}=0, ~~~\frac{4}{3}+\frac{\partial\beta_{12}}{\partial u}+\beta_{22}\frac{\partial\beta_{22}}{\partial u}=0,$$
which have no common solution, thus $\beta_{33}=0$. Then we consider two cases $\beta_{21}\neq0$ and $\beta_{21}=0$.  For the former case, substituting the relations (\ref{equa-3}), (\ref{equa-1}) and (\ref{equa-2}) into Eq.(\ref{kdv-inva-kp1}), then extracting coefficient of $u_3v_3u_1^3$ yields $-3(c_1-u)^5$ which is not zero and thus no solutions exist for this case.

Consider the latter case $\beta_{21}=0$. Then  $\beta_{11}=0,\beta_{01}=-1,\beta_{32}=\beta_{22}$, and four crucial equations of the determining system are
\begin{eqnarray}\label{equa-4}
&&\no 3\frac{\partial\beta_{12}}{\partial u}+2=0,\\
&&\no 3\frac{\partial^2\beta_{02}}{\partial u^2}-4\frac{\partial\beta_{22}}{\partial u}=0,\\
&& \no\beta_{22}\left(1+3\beta_{22}\frac{\partial\beta_{22}}{\partial u}\right)=0,\\
&& 3\beta_{02}\left(1+\frac{\partial\beta_{12}}{\partial u}+2\beta_{22}\frac{\partial\beta_{22}}{\partial u}\right)-\beta_{12}\beta_{22}=0.
\end{eqnarray}

The first equation gives $\beta_{12}=-2u/3+c_1$ and the third equation gives $\beta_{22}=\pm\sqrt{2(3c_1-u)/3}$ or $0$ respectively. If $\beta_{22}=\pm\sqrt{2(3c_1-u)/3}$, then substituting them into the last equation we obtain
$\beta_{02}=\left(3 c_1-2 u\right) \sqrt{2(3 c_1-u)/27}$.
With such results, the second equation becomes $$3\frac{\partial^2\beta_{02}}{\partial u^2}-4\frac{\partial\beta_{22}}{\partial u}=\frac{\sqrt{\frac{3}{2}} \left(2 u-3 c_1\right)}{2 \left(3 c_1-u\right){}^{\frac{3}{2}}}\neq0,$$
thus $\beta_{22}=0$ and follows by $\beta_{02}=0$ from the last equation in system (\ref{equa-4}). Thus we obtain the invariant manifold with $p=4$ as (\ref{in-p=4}).
The proof ends. $\hfill{} \Box$

\subsection{The mKdV equation}
Another example is the modified KdV equation (mKdV)
 \begin{equation}\label{pot-kdv}
w_t+\frac{1}{6}w^2w_x-w_{xxx}=0,
\end{equation}
which is connected with Eq.(\ref{eqn}) by $u=-w^2/6+w_x$. The linearized equation of Eq.(\ref{pot-kdv}) is
 \begin{equation}\label{pot-kdv-linear}
v_t+\frac{1}{3}ww_xv+\frac{1}{6}w^2v_x-v_{xxx}=0.
\end{equation}

We assume the linear invariant manifold takes the form
 \begin{equation}\label{kdv-ivar}
  H^p=v_p-\sum_{j=0}^{p-1}\alpha_jv_j=0,
\end{equation}
where $\alpha_j=\alpha_j(u,u_x,u_{xx},u_{xxx})$ and $p\leq 7$ by Theorem \ref{the-2}. The invariant manifold with $p=2$ and $p=3$ have been considered in \cite{hab-2016} via the potential KdV equation. Thus here by similar and tedious computations as in Section 3.1, we find  two invariant manifolds with $p=4$ of Eq.(\ref{pot-kdv-linear}) given by
\begin{eqnarray}\label{kdv-pot-1}
&&\no H^4_1=v_4-\,v_3\left[\frac{u_1 u^2+3 u_1^2-9 u_3}{3 \left(u u_1-3 u_2\right)}\right]\\
 && \no\hspace{1.5cm} -\,v_2 \left[\lambda+\frac{1}{9}u^2+\frac{3u_3 u-u_2 u^2-3u_1
   u_2}{3 \left(u u_1-3 u_2\right)}\right]\\
 && \no\hspace{1.5cm} +\,v_1 \left[\frac{9 \lambda u_1 u^2+27 \lambda u_1^2-81 \lambda u_3+u_1 u^4+3u_1^2 u^2-9 u_3
   u^2}{27 \left(u u_1-3 u_2\right)}-\frac{5}{9}uu_1\right]\\
 && \no\hspace{1.5cm} -\frac{v }{27 \left(u u_1-3 u_2\right)}\big[9 \lambda u_2 u^2-27 \lambda u_3 u+27 \lambda u_1 u_2+u_2 u^4-3 u_1^2
   u^3\\
   && \hspace{4cm}-3 u_3 u^3 +15 u_1 u_2 u^2-36 u_2^2 u+27 u_1 u_3 u-27
   u_1^2 u_2\big]=0,
\end{eqnarray}
and \begin{eqnarray}\label{kdv-pot-2}
&&\no H^4_2=v_4+v_3\left[\frac{u_1 u^2-3 u_1^2-9 u_3}{3 \left(u u_1+3 u_2\right)}\right]\\
 && \no\hspace{1.5cm} - \,v_2 \left[\lambda+\frac{1}{9}u^2+\frac{u_2 u^2+3 u_3 u-3 u_1
   u_2}{3\left(u u_1+3 u_2\right)}\right]\\
 && \no\hspace{1.5cm} -\, v_1\left[\frac{9 \lambda u_1 u^2-27 \lambda u_1^2-81 \lambda u_3+u_1 u^4-3 u_1^2 u^2-9 u_3
   u^2}{27 \left(u u_1+3 u_2\right)}+ \frac{5}{9}uu_1\right]\\
 && \no\hspace{1.5cm} + \frac{v}{27 \left(u u_1+3 u_2\right)} \big[9 \lambda u_2 u^2+27 \lambda u_3 u-27 \lambda u_1 u_2+u_2 u^4-3 u_1^2
   u^3\\
   &&\hspace{4cm}+3 u_3 u^3-15 u_1 u_2 u^2-36 u_2^2 u+27 u_1 u_3 u-27
   u_1^2 u_2\big]=0.
\end{eqnarray}
where $\lambda$ is a constant parameter.

With similar proof of Proposition \ref{prop-1} for KdV equation, the existence of invariant manifold with the form (\ref{kdv-ivar}) for Eq.(\ref{pot-kdv-linear}) is summarized as follows.
\begin{prop}
For $p=4$, the invariant manifolds of Eq.(\ref{pot-kdv-linear}) are given by Eq.(\ref{kdv-pot-1}) and  Eq.(\ref{kdv-pot-2}).
For $5\leq p\leq7$, there are no invariant manifolds of the form (\ref{kdv-ivar}) for Eq.(\ref{pot-kdv-linear}).
\end{prop}

Moreover, with the computer algebra software, it is not difficult to verify that $X_i=H^4_i\partial_v\,(i=1,2)$ are two generalized conditional symmetries of the system composed of Eqs.(\ref{pot-kdv}) and (\ref{pot-kdv-linear}). The two invariant manifolds (\ref{kdv-pot-1}) and (\ref{kdv-pot-2}) together with Eq.(\ref{pot-kdv-linear}) form two Lax pairs of Eq.(\ref{pot-kdv}).
\subsection{The Sharma-Tasso-Olver equation}
The Sharma-Tasso-Olver equation takes the form
\begin{eqnarray}\label{sto}
&& u_t+3\gamma u^2u_x+3\gamma u_x^2+3\gamma uu_{xx}+\gamma u_{xxx}=0,
\end{eqnarray}
which is a prominent double nonlinear dispersive model and contains the linear dispersive term $\gamma u_{xxx}$ and the double
nonlinear terms $\gamma (u^3)_x$ and $\gamma (u^2)_{xx}$ \cite{olv-1977}.
Taking the transformation $u=w_x,\widehat{t}=\gamma t$ into Eq.(\ref{sto}) and integrating once we obtain
\begin{eqnarray}\label{sto-pot}
&&w_t+ w^3_x+3 w_xw_{xx}+ w_{xxx}=0,
\end{eqnarray}
where the integrated constant is assumed to be zero and the hat ``~${\widehat{}}$~'' on $t$ is omitted. The linearized equation of Eq.(\ref{sto-pot}) is
\begin{eqnarray}\label{sto-potlinear}
&&v_t+3w_x^2v_x+3v_xw_{xx}+3w_xv_{xx}+v_{xxx}=0.
\end{eqnarray}

We search for the invariant manifold of Eq.(\ref{sto-potlinear}) with the form (\ref{kdv-ivar}), then by Theorem \ref{the-2} the order $p$ is bounded by $p\leq 7$.
Following similar procedure for KdV equation, we obtain three invariant manifolds $H^2_i=0\,(i=1,2,3)$ given by
\begin{eqnarray}
&&\no H^2_1=v_2-v_1\left(\lambda-w_1+\frac{2w_1w_2+w_3}{w_1^2+w_2}\right)\\
 && \no\hspace{1.5cm} -\frac{v}{w_1^2+w_2} \left[w_2 \left(w_1^2-2 \lambda w_1-w_2\right)+w_3 \left(w_1-\lambda\right)\right],\\
&&\no H^2_2=v_2+v_1\left[w_1 +\frac{w_2 \left(2w_1-\lambda\right)+w_3}{w_1 \left(\lambda-w_1\right)-w_2+\lambda}\right]\\
   &&\no\hspace{1.5cm}+\frac{v}{w_1 \left(\lambda-w_1\right)-w_2+\lambda}\left[w_2\left(w_1^2-w_2+\lambda\right)+w_1 w_3\right],\\
&&\no H^2_3=v_2+\frac{v_1}{\lambda  w_1^2+w_2 \left(\lambda -e^{w}\right)}\left(\lambda w_1^3-\lambda  w_1 w_2-\lambda w_3+e^{w}w_3\right)\\
   &&\hspace{1.5cm}+\frac{\lambda v}{\lambda  w_1^2+w_2 \left(\lambda -e^{w}\right)}\left(w_2^2-w_1^2 w_2- w_1 w_3\right),
\end{eqnarray}
and seven invariant manifolds $H^3_j=0\,(j=1,\dots,7)$ expressed by
\begin{eqnarray}
&&\no H^3_1=v_3+v_2\frac{2 w_1^2 \left(\lambda-w_1\right)-\lambda w_2+w_3}{w_2+w_1^2-\lambda w_1} +v_1\frac{\lambda w_1^3+w_1 \left(\lambda w_2+2 w_3\right)-w_1^4-3 w_2^2}{w_2+w_1^2-\lambda w_1}\\
 && \no\hspace{1.5cm} +v\frac{\lambda\left(w_1^2-w_2\right) w_2+\lambda w_1 w_3}{w_2+w_1^2-\lambda w_1},\\
   &&\no H^3_2=v_3+v_2\frac{w_1 \left(w_1+\lambda\right) \left(2 w_1+\lambda\right)-w_3}{w_1
   \left(w_1+\lambda\right)+w_2}\\
 && \no\hspace{1.5cm} +v_1\frac{2 \lambda w_1^3+\lambda^2 (w_1^2-w_2)-w_3 \left(2 w_1+\lambda\right)+w_1^4+3
   w_2^2}{w_1 \left(w_1+\lambda\right)+w_2},\\
   &&\no H^3_3=v_3+v_2\frac{2 w_1^3-w_3}{w_1^2+w_2}+v_1\frac{\lambda w_1^2+w_2 \left(3 w_2+\lambda\right)+w_1^4-2 w_3
   w_1}{w_1^2+w_2}-v\frac{\lambda \left(2 w_1 w_2+w_3\right)}{w_1^2+w_2},\\
  &&\no H^3_4=v_3-v_2\frac{2 \lambda w_1-2 w_1^3+w_3}{w_2+ w_1^2-\lambda} -v_1\frac{4 \lambda w_2+2 \lambda w_1^2-w_1^4+2 w_3 w_1-3
   w_2^2-\lambda^2}{w_2+ w_1^2-\lambda},\\
&&\no H^3_5=v_3+\frac{v_2}{\Delta}\left[3 \lambda w_1+e^{w} \left(2 w_1^3-w_3\right)\right]+\frac{v_1}{\Delta}\left[3 \lambda \left(w_1^2+w_2\right)+e^{w} \left(w_1^4-2 w_3 w_1+3 w_2^2\right)\right]\\
   &&\no\hspace{1.5cm}+\frac{v}{\Delta}\left[\lambda \left(w_1^3+3 w_2 w_1+w_3\right)\right],\\
&&\no H^3_6=v_3+v_2 \frac{w_3 \left(\lambda e^{w}-1\right)-\lambda e^{w} w_2 w_1+2 w_1^3}{w_1^2-w_2
   \left(\lambda e^{w}-1\right)}+v_1\frac{w_3 w_1 \left(\lambda e^{w}-2\right)-w_2^2 \left(2 \lambda
   e^{w}-3\right)+w_1^4}{w_1^2-w_2 \left(\lambda e^{w}-1\right)},\\
&& H^3_7=v_3+v_2 \frac{2 w_1^3-w_3}{w_1^2+w_2}+v_1\frac{w_1^4-2 w_3 w_1+3 w_2^2}{w_1^2+w_2},
\end{eqnarray}
where $\Delta=e^{w} w_1^2+e^{w} w_2+c_2$ and $\lambda$ is a constant parameter.

It is easy to show that $X_i=H^2_i\partial_v\,(i=1,2,3)$ and $\widetilde{X}_j=H^3_j\partial_v\,(j=1,\dots,7)$ are  generalized conditional symmetries of Eqs.(\ref{sto-pot}) and (\ref{sto-potlinear}), and then Eq.(\ref{sto-potlinear}) together with $H^2_i=0\,(i=1,2,3)$ and $H^3_j=0\,(j=1,\dots,6)$ constitute nine Lax pairs of Eq.(\ref{sto-pot}) since $H^3_7=0$ does not involve constant parameter $\lambda$ and therefore does not generate any true Lax pair. Note that we omit the cases for $4\leq p\leq7$ since they can be studied with similar techniques and contain labor-consuming computations.

\section{Conclusion}
We formulate the method in \cite{hab-2016,hab-2017} in the context of generalized conditional symmetry and give an upper order bound of the derivatives appearing in the invariant manifold, which provides a theoretical basis for the complete classification of the given form invariant manifold and then for the Lax pair. We illustrate the results by three examples.
\section*{Acknowledgements}
This paper is supported by the National Natural Science Foundation of China (No.11671014).

\end{document}